# Girlhood Feminism as Soft Resistance: Affective Counterpublics and Algorithmic Negotiation on RedNote

**Meng, Liang; Xiaoyue, Zhang; Linqi, Ye**

## Abstract

This article explores how Chinese female users tactically mobilise platform features and hashtag practices to construct vernacular forms and an exclusive space of feminist resistance under algorithmic and cultural constraints. Focusing on the reappropriation of the hashtag #Baby Supplementary Food (#BSF), a female-dominated lifestyle app with over 300 million users, we analyse how users create a female-centered counterpublic through self-infantilisation, algorithmic play, and aesthetic withdrawal. Using the Computer-Assisted Learning and Measurement (CALM) framework, we analysed 1,580 posts and propose the concept of "girlhood feminism": an affective, culturally grounded form of soft resistance that refuses patriarchal life scripts without seeking direct confrontation or visibility. Rather than challenging censorship and misogyny directly, users rework platform affordances and domestic idioms to carve out emotional and symbolic spaces of dissent. Situated within the broader dynamics of East Asia's compressed modernity, this essay challenges liberal feminist paradigms grounded in confrontation and transparency. It advances a regionally grounded framework for understanding how gendered publics are navigated, negotiated, and quietly reimagined in algorithmically governed spaces.

## Introduction

On RedNote, a female-dominated lifestyle platform with over 300 million users, a seemingly mundane parenting hashtag—#Baby Supplementary Food (#BSF)—has taken on unexpected political and affective significance. Originally intended for sharing baby food recipes, #BSF has been widely repurposed by female users to tag personal posts such as selfies, pet photos, and emotional reflections. This reappropriation operates as a subtle tactic of algorithmic navigation: based on a shared belief that parenting-related hashtags are less likely to be promoted to male audiences, users employ #BSF to shield intimate content from unwanted exposure (Cui, 2023) . In doing so, the hashtag functions as a form of algorithmic camouflage—transforming a domestic, non-political tag into a gendered filter that facilitates controlled visibility, emotional safety, and quiet withdrawal from public scrutiny.

This tactic，in this way, creates a women-centered counterpublic—an alternative discursive space shaped through cultural re-signification and platform-savvy use. By appropriating #BSF, users sidestep the male gaze while fostering intimacy, solidarity, and shared affect. Hashtags here act not just as tools of categorisation (Losh, 2019; Kreiss & McGregor, 2024), but as feminist techniques of spatialisation, forming safe, networked counterpublics (Fraser, 1992; Jackson et al., 2020; Kanai & McGrane, 2020). These practices reflect what Bucher (2016) calls algorithmic imaginaries—vernacular understandings of algorithmic systems tactically repurposed for cultural ends (Schulz, 2022; Gran et al., 2020; DeVito et al., 2018).

This study uses the #BSF hashtag on RedNote as a case to explore how such resistance is practiced, felt, and negotiated. It is based on traditional and computational grounded theory (CALM framework) to analyse 1,580 #BSF posts collected from Rednotes, using GSDMM clustering, open coding, and fine-tuned BERT classification to identify 5 significant thematic categories reflecting feminist expressions in a Chinese digital context.

This article introduces *girlhood feminism* as a form of vernacular resistance on Chinese digital platforms, enacted through affective withdrawal, algorithmic play, and strategic self-infantilisation. While existing scholarship often frames self-infantilisation as a retreat from neoliberal adulthood—marked by pressures of self-regulation and economic individuation—in China, it signals resistance to a different set of obligations: filial duty, marriage, and motherhood, embedded in Confucian kinship structures. In both contexts, girlhood operates as a refusal of normative life scripts. Yet Chinese expressions of girlhood assert emotional and lifestyle autonomy against patriarchal family roles. By tracing this shift, we argue that girlhood feminism reclaims infantilisation not as submission, but as a culturally specific strategy of soft resistance.

Our analysis identifies two core dynamics within this formation. First, self-infantilisation functions not as immaturity but as affective disidentification—a coded way of saying: *I am not a mother/wife, and I will not become one—at least not within the feudal family structure*. Second, girlhood feminism gives rise to a non-institutional, emotionally mediated counterpublic, grounded in aesthetic improvisation and everyday acts of soft resistance. Meanwhile, unlike more consolidated cultural formations such as Japanese shōjo culture or liberal feminism (Ryan, 1992), girlhood feminism in China is fragmented, context-specific, and vernacular—emerging from the rhythms of daily life as a particular soft resistance.

This conceptual intervention contributes to feminist digital media studies by highlighting how affective and non-confrontational tactics function as legitimate modes of resistance, particularly how the identity of *girlhood* acquires distinct political significance across cultural and political contexts. It also contributes to platform governance and algorithmic publics by illustrating how users tactically navigate algorithmic systems to manage visibility and audience exposure. Finally, it extends East Asian media and gender studies by theorizing a regionally grounded feminist formation shaped by the contradictions of compressed modernity. This approach challenges liberal models of feminist resistance that often overlook the informal, emotionally mediated, and locally embedded strategies emerging in contexts like China (Peng, 2018; Yang & Guo, 2023). By centering platform-based vernacular feminism, this study expands the analytical vocabulary through which gendered resistance in algorithmic spaces can be understood.

**RedNote Counterpublics: Affective Hashtag Practices and Algorithmic Imaginaries**

In digital environments, hashtags have become key tools for marginalized groups to construct counterpublics—alternative discursive spaces that resist dominant narratives and enable collective identity formation. Building on Fraser's (1992) critique of Habermas's unitary public sphere, counterpublic theory highlights how subordinated groups circulate oppositional discourse outside bourgeois norms.

Affection plays a central role in sustaining these counterpublics. As Warner (2002) argues, a public is not solely defined by the public interests of its participants but also serves as a space for the circulation of intimate relations and affects—including embodied experience, erotic expression, and practices of care (Warner, 2002). In this sense, members of counterpublics are able to share private aspects of their lives with others. Through these practices, counterpublics not only contest the norms of dominant discourse, but also blur the boundaries between public and private, reclaiming personal and intimate expression as politically significant forms of public engagement.

This becomes especially relevant when considering female counterpublics, which Berlant (2008) terms intimate publics. As she argues, women's culture is cultivated through "a porous, affective scene of identification" (p. viii). These intimate publics are bound together by shared emotional experiences and a broadly lived social history rooted in reproductive labor and mutual legibility—spaces where personal feelings are transformed into collective claims. In this sense, affective sharing becomes one of the most important bonding mechanisms within female counterpublics. Moreover, in feminist counterpublics, emotional resonance is not merely a tool for mobilization; it becomes a political form in itself, shaping how marginalized users navigate issues of visibility, recognition, and voice.

Yet, in the digital media context, these counterpublics are also deeply shaped by platform infrastructures, especially the use of hashtags. Beyond direct articulation, users engage in strategic practices such as amplification, evasion, and hijacking (Treré & Bonini, 2022). These strategies are grounded in what Bucher (2016) calls algorithmic imaginaries—users' evolving, experience-based understandings of algorithmic systems, which inform how platforms can be tactically reconfigured. Through everyday interactions (Schellewald, 2022), such imaginaries transform platform mechanics into vernacular tools of digital folklore (Jones, 2023), allowing hashtags to carry not just content, but suppressed desires, needs, and political imaginaries.

Within this broader landscape, *#BSF* represents a particularly salient instantiation of algorithmic imaginary in practice. On RedNote, a platform largely populated by female users, the appropriation of *#BSF* reflects a shared belief that parenting-related hashtags are less likely to trigger algorithmic exposure to male audiences. Mobilizing this imagined algorithmic affordance, *#BSF* operates as a vernacular tactic of audience management, enabling users to regulate the visibility of intimate or sensitive content. Through this everyday maneuver, a mundane parenting tag is re-signified into a gendered shield of soft resistance, producing affective zones of intimacy, protection, and quiet refusal within algorithmically governed publics.

**Between State, Market, and Misogyny: Feminist Discourse under Compressed Modernity**

China's feminist discourse unfolds within a contested terrain shaped by state ideology, market reforms, digital infrastructures, and resurgent misogyny. Confucian traditions, socialist legacies, and neoliberal logics intersect, generating competing demands from state, market, and cultural norms. To grasp these tensions, we draw on the framework of compressed modernity—originally developed in relation to South Korea, where rapid, overlapping transformations have restructured political, economic, and cultural life (Chang, 2010). Scholars have since extended this concept to other East Asian contexts marked by similar histories of Confucian patriarchy and neoliberal capitalism (Ma, 2012). In China, compressed modernity manifests as a hybrid condition combining state-regulated capitalism, authoritarian governance, and truncated individualisation (Lu & Holbrook, 2014; Beck & Beck-Gernsheim, 2010), creating a volatile discursive space that feminist counterpublics must tactically navigate.

Under conditions of compressed modernity, the state has played a central role in shaping feminist discourse in China, alternating between promoting women's public participation and reinforcing domestic ideals in line with shifting political and economic goals. In the Maoist era, women's liberation aligned with revolutionary labour (Yin, 2021). Since the 1980s, however, market reforms and a revival of Confucian values have repositioned women as caregivers within the extended kinship family (Fei, 1992), supporting state visions of harmony and economic restructuring (Leung, 2003; Gaetano, 2014).

Here, "family" largely reflects Confucian patrilineal systems, not the nuclear unit. Women are expected to perform intergenerational care within patriarchal structures that emphasise filial and collective obligations. This ideological shift has been accompanied by structural change: the retreat of state welfare and redistribution of reproductive labour to households. Across East Asia, what Ochiai (2011) terms the "failure of familialism" highlights how families, especially women, are overburdened with care but lack systemic support. In China, this manifests through gendered roles embedded in both market and Confucian familial structures (Berik et al., 2007).

Simultaneously, China's embrace of neoliberal wave gave rise to a consumerist model of femininity. Feminist rhetoric was appropriated as a marketing tool, offering the illusion of empowerment through consumption rather than advocating for real structural change (Peng, 2018; Wang & Zhang, 2023). Women were encouraged to cultivate self-worth through beauty, lifestyle, and emotional labour—performing normative "feminine traits" under the banner of choice. RedNote itself also exemplifies this logic, having evolved from shopping-oriented communities into discursive spaces shaped by neoliberal femininity. These dynamics have increasingly depoliticised feminist discourse, reframing empowerment as individual self-optimization rather than collective critique.

The rise of internet platforms has paradoxically amplified both feminist discourse and its suppression, rendering feminist expression increasingly fragmented and tactical. While digital spaces have enabled greater visibility for feminist discussions despite strict state controls (Banet-Weiser, 2018; Ye & Huang, 2022; Tan & Liu, 2024), they are far from neutral. Under the state's "harmonious society" discourse, feminism is framed as a threat to social stability—justifying asymmetrical platform governance, where feminist content is swiftly censored while misogynistic speech remains largely untouched (Jing, 2020; Huang, 2021; Liao, 2023). This alignment sustains what Ye and Huang (2022) and Tan and Liu (2024) describe as "platformised misogyny". Feminist activists face content takedowns, defamation, and algorithmic erasure. In response, feminist discourse turns to subtle, coded, and affectively resonant forms of soft resistance—strategies that sustain critical engagement and collective solidarity while navigating censorship.

Meanwhile, digital platforms are embedded in masculine speech norms that marginalise feminist voices (Ging, Siapera, & Chemaly, 2019), fueling a wave of online misogyny. Feminist activists face coordinated defamation and are labeled with slurs such as "rural feminists" (田园女权), "feminist bitches" (女权婊), and "feminist dogs" (女权狗) (Wang & Zhang, 2022). As Yang and Guo (2023) argue, such backlash often stems from a misrecognition of feminist goals—mistakenly assuming Chinese feminism mirrors Western agendas, while ignoring local conditions that preclude such demands. This disjuncture reflects compressed modernity, where rapid modernisation outpaces the development of supportive gender structures. As a result, Chinese feminism is both hyper-visible and structurally misunderstood—a contradiction rooted in the region's uneven modernisation. Within this context, feminist activism becomes increasingly fragmented, tactical, and emotionally coded.

Against this backdrop of compressed modernity and systemic gendered constraints, the hashtag #BSF on RedNote exemplifies a distinctly Chinese mode of feminist resistance—one that exceeds liberal feminist paradigms. Rather than relying on visibility or institutional confrontation, it operates through strategic ambiguity, emotional coding, and algorithmic play. This study adopts a grounded theory approach to trace these emergent formations within China's gendered platform ecology, centring affective tactics and vernacular negotiation.

**Methodology**

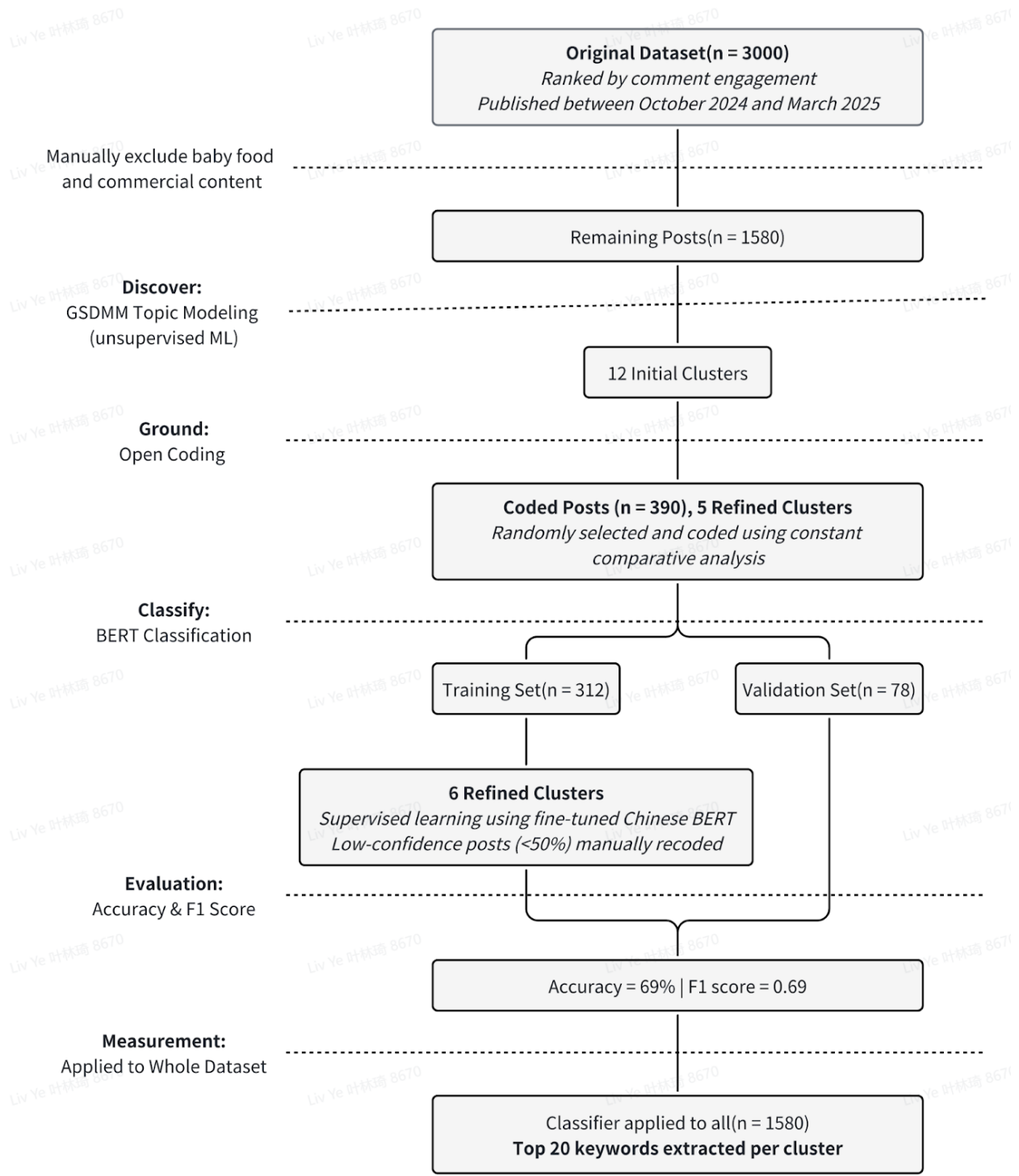

Figure 1: Analytical Workflow Integrating CALM Framework and Grounded Theory

This study adopts the Computer-Assisted Learning and Measurement (CALM) framework (Carlsen & Ralund, 2022), extending grounded theory through computational techniques (Strauss & Corbin, 1998; Nelson, 2020). The analysis was conducted by three Chinese female researchers, each with over five years of experience on Rednotes, ensuring familiarity with the platform's culture and vernacular.

We collected 3,000 top-ranked #BSF posts (Oct 2024–Mar 2025) from Rednotes' official data portal. After manually filtering out baby food tutorials and commercial content, 1,580 valid posts remained. Initial theme discovery was conducted using unsupervised clustering (GSDMM), followed by open coding of 390 randomly selected posts. Through constant comparative analysis, five core thematic clusters were refined. Supervised classification with a fine-tuned Chinese BERT model expanded this to six clusters. Posts with low prediction confidence (<50%) were manually recoded. Model performance was moderate (accuracy = 69%, F1 = 0.69). The validated classifier was applied to the full dataset, and the top 20 keywords from each cluster were extracted to support word cloud visualization. One cluster focused on cosmetic giveaways with pronounced neoliberal features, and is excluded from detailed analysis in later sections.

**Findings**

**Cluster 1 Baby Food and Bottled Selves: Food as Social Linkage and Self-Infantilization**

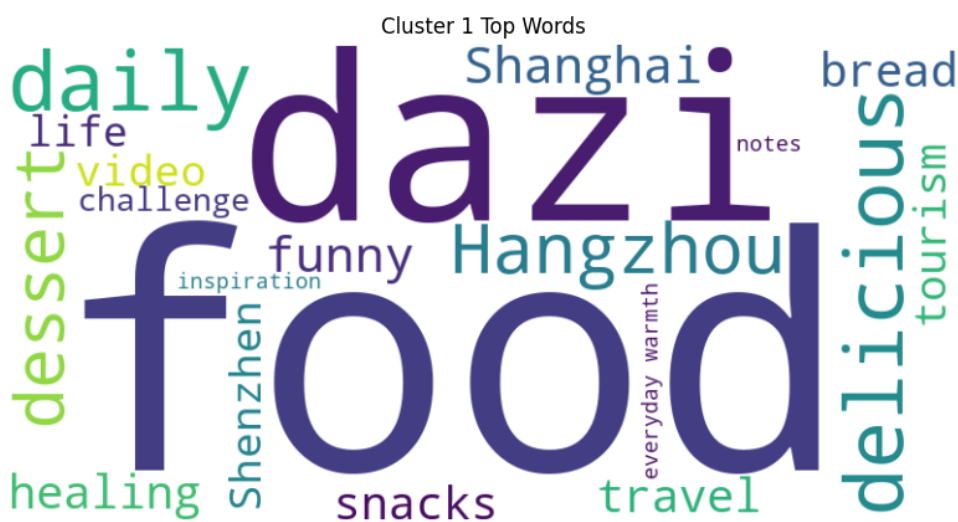

Figure 2: Word Cloud of High-Frequency Keywords in Cluster 1

Food-related posts constitute the largest thematic cluster within #BSF (690). While ostensibly focused on meal sharing or dietary suggestions, these posts function as emotionally charged narratives, where food becomes a medium for affective expression. Central to this is the tactic of self-infantilization, through which users adopt a "baby" persona to articulate vulnerability and care. In this process, the parenting-associated hashtag #BSF is re-signified as a vernacular grammar of intimacy and emotional retreat.

In this cluster, word frequency analysis highlights "food" (373) as the dominant term, followed by "dazi" (153), "daily life" (97), "delicious" (91), "dessert" (80), and "bread" (55). These keywords suggest a threefold semantic structure: affect, everydayness, and social bonding. The prevalence of "dazi" (slang for "eating companion") reflects a desire for para-social intimacy, as seen in phrases like "@my dazi" or "Who will be my dazi tomorrow?" Terms like "daily," "life," and "video" evoke the platform's fragmented, narrative-driven formats, aligning closely with the visual–narrative formats of short videos and photo captions. These formats foster a cycle of documenting, sharing, and emotional feedback. Meanwhile, words such as "delicious," "dessert," and "healing" point to food's emotional function as comfort, attachment, and affective release.

This affective logic is evident in the dual use of food personification and infantile tone. For instance, in phrases like "I'm addicted to this cream! It holds up my whole world!" or "The cream is heaven and earth for me," food is not just a culinary item but a surrogate for emotional release. The other examples include that one user refers to a bowl of spinach protein rice as "Starbucks-level crack at my school," claiming, "I'd die if I couldn't eat this anymore," reframing food as a form of emotional dependence. Another writes, "This blue-skin girl is just too cute! I had to make an onigiri version—perfect for Halloween ghost bento," folding food into a layered expression of seasonal aesthetics, cuteness, and creativity. In this vernacular system, food acquires symbolic weight beyond its physical function. These narratives do not dwell on culinary processes; instead, they center around the moods, emotional needs, and therapeutic possibilities that food can metaphorically fulfill. Similarly, posts like "Try it now or @ your dazi to make it for you~" evoke a childlike, playful tone of intimacy. Repeated use of phrases such as "I'm crying-crying," "I'm fainting from happiness," and baby-talk expressions solidify a recognizable emotive-aesthetic grammar that translates emotional input into legible content output.

Linguistically, many of these posts adopt a first-person register—"I," "this baby," or "we babies"—while employing affectionate, exaggerated, or melodramatic tones. This self-positioning as someone who seeks care mimics infant speech patterns and evokes a soft, lovable persona. Such stylistic "regression" participates in an emotional format and a standardised tone. Posts often include emojis (e.g., 🥺, 😊), reduplication ("soft-fufu," "yummy-yummy"), and romantic or addictive metaphors ("feels like falling in love," "I can never quit it"), all of which forge a tight link between food and emotional states.

**Cluster 2 Pets as "Babies": The Parody of Motherhood**

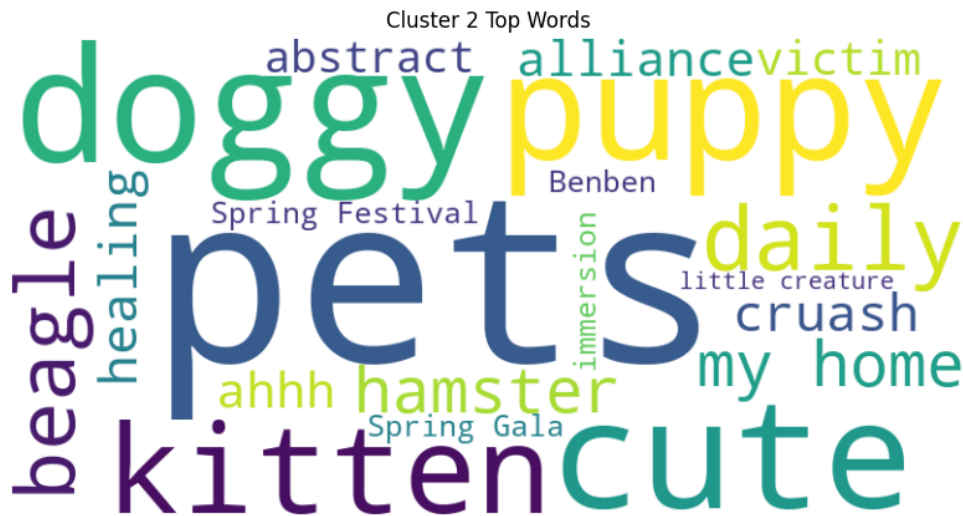

Figure 3: Word Cloud of High-Frequency Keywords in Cluster 2

The second thematic cluster focuses on pet-related content. While these posts ostensibly document the daily lives of pets, cute moments, or feeding routines, they in fact stage a stylised enactment and parody of maternal care. In contrast to the first cluster—where users position themselves as "babies" in need of care—this cluster features users adopting the opposite role of the caregiver. Through the motif of "feeding" pets, users rehearse a maternal identity that is playful, affectively charged, and emotionally expressive.

Word frequency analysis reveals that the most frequent terms include "pet" (70), "doggy" (37), "cute" (33), "puppy" (32), "kitten" (27), "daily" (26), "beagle" (22), and "hamster" (18). These words constitute a highly affective and domestic semantic field. On one hand, possessive expressions such as "my family dog" or "my family cat" position pets as part of a familial structure, invoking the idiom of kinship. On the other hand, the high frequency of the word "daily" underscores the ordinary, repetitive nature of caregiving practices—feeding, cleaning, playing, and monitoring health—situated firmly within the rhythm of home life. Moreover, the presence of holiday-related terms such as "Spring Festival" and "Spring Gala" reveals how pets are further embedded into the emotional rituals of the household, especially during culturally significant moments. These elements together naturalise the portrayal of users as "mothers," and pets as emotionally expressive, dependent "babies."

Meanwhile, the emotion embedded in this content is often presented through a tone of melodramatic humour or exaggerated self-deprecation. Phrases like "ahhh," "who can understand," or "I'm going crazy" are not cries for help, but rather stylistic devices to translate daily caregiving routines into legible emotional performance. At the same time, terms such as "healing," "cruash" (a playful rendering of "crush"), and "abstract" further index a vernacular of cuteness and affect, blending sincere emotional attachment with layers of irony and platform-specific humour.

Additionally, the frequent naming and characterisation of specific pets—such as "Benben" (which is a common name for pets, which means "silly")—suggests that pets are not generic stand-ins for companionship, but highly personalised and narratively elaborated figures. These animals often appear in serialised storytelling formats, complete with personalities, backstories, and emotional arcs, echoing the structure of parenting diaries or miniature domestic dramas.

**Cluster 3  The Weight of Care: Familial Labor, Education, and Intergenerational Strain**

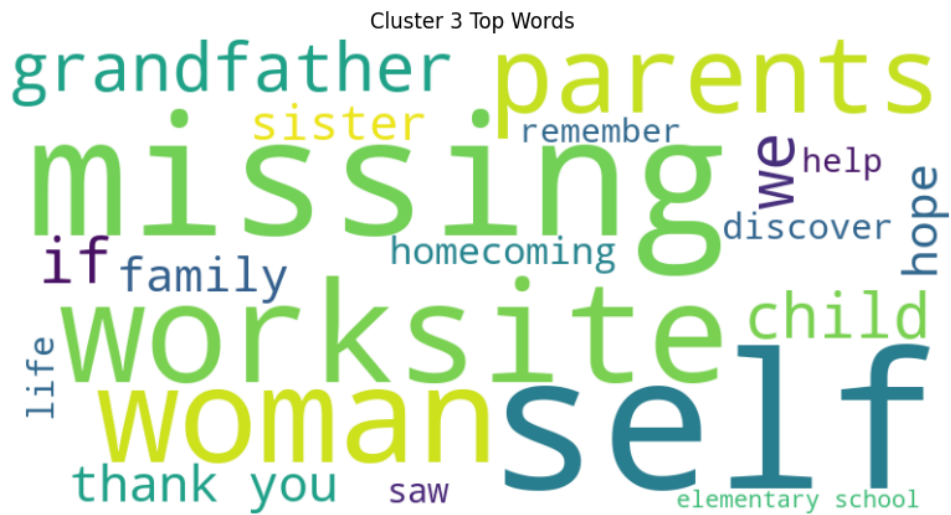

Figure 4: Word Cloud of High-Frequency Keywords in Cluster 3

This cluster foregrounds the burdens faced by women within intergenerational family structures shaped by Confucian and feudal kinship systems. High-frequency terms like "self" (151), "woman" (81), "parents" (77), and "grandfather" (66) reflect a discourse saturated with extended family roles and domestic tension. Users reflect critically on the gendered division of labour, particularly the undervaluation of women's caregiving, emotional labour, and physical sacrifices. One user recalls watching her cousin's wife arrive at dinner carrying newborn twins—exhausted, unassisted, and unacknowledged—concluding: "They trade physical labour for the illusion of a stable family." In this context, male financial contributions are often praised, while women's affective and bodily investments are rendered invisible.

This gendered double standard is not just observed, criticised but resisted. Posts critique both the romanticization of motherhood and the emotional depletion it entails. In one striking example, a user writes: "Reminder to self: never become a woman like my mom," which shows a resistance to traditional female life scripts. Across this cluster, women collectively question whether familial devotion is worth the personal cost.

Amid critiques of familial obligation, education emerges as a feminist counter-strategy—a path toward self-reliance and financial independence. Many users frame education as liberation from patriarchal marriage, often with humorous resolve: "We eat readings, sleep under Chicago, hang APA on the wall..." One writes: "I just want to grow up, get a job that supports me, and rent a tiny apartment." In this account, financial independence is seen as a way for girls to avoid entering patriarchal marriages, with education regarded as the most effective means to achieve it.

The final dimension of this cluster highlights the trafficking of women into forced marriages in rural areas, where their reproductive abilities are exploited to sustain family lineages. The keyword "missing" appears 116 times, and "homecoming" 51 times, often in relation to lost or abducted girls. Users post calls for help in locating sisters or daughters trafficked decades earlier: "Born November 26, 1997... she's been missing for 20 years. She is my sister. Please help share this post." In this sense, these posts reactivate the collective memory of women's historical trauma and mobilize mutual aid networks as users come together to help locate missing family members.

**Cluster 4 Girls Looking at Girls: The Aesthetics of Everyday Becoming**

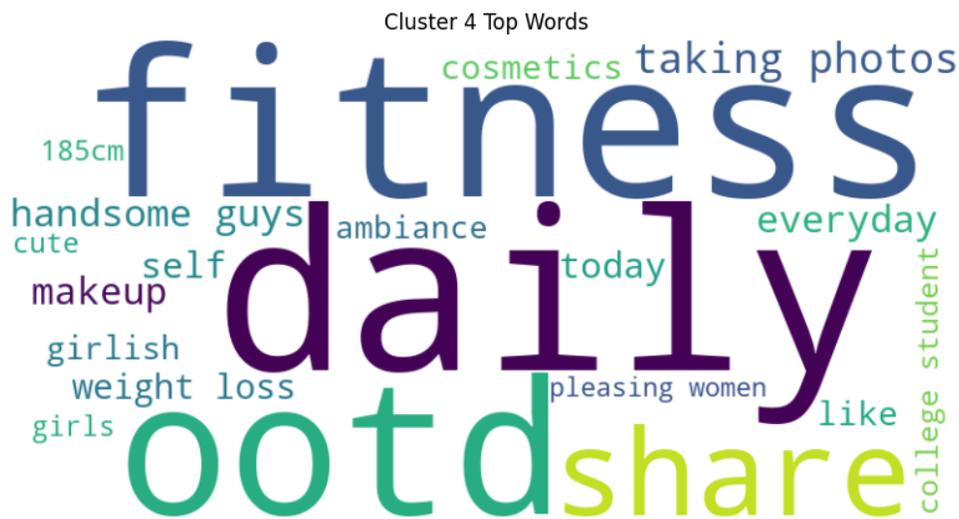

Figure 5: Word Cloud of High-Frequency Keywords in Cluster 4

Cluster 4 explores female subjectivity through body, mood, and daily routines, articulated from a female-first perspective (e.g., "self," 17). Keywords such as "daily" (29), "today" (17), "taking photos" (23), and "everyday" (19) reflect a fragmented, embodied vernacular. Aesthetic labour is central: "fitness" (28),

"makeup" (18), "cosmetics" (15), "OOTD" (25), and "weight loss" (16) highlight body styling as expressive practice. Crucially, this is not self-objectification but a reorientation of value. The term "pleasing women" (媚女, 10) signals a shift from attracting men to impressing and affiliating with other women—exemplified by posts like: "Today I'm serving alt-girl realness—my wig and makeup combo is so good I impressed myself."

The female gaze here is also playful, ironic, and reflexive. Linguistically, this cluster also reclaims expressivity through exaggerated self-parody. Anti-disciplinary humour—e.g., toilet jokes, mock barking—challenges expectations of feminine restraint: "Since I was a child, I've loved barking like a dog… Now I just desperately want a place to express myself and bark thoroughly."

Meanwhile, users blur the line between subject and object—styling, watching, and interpreting themselves and others. Male bodies appear too, but not as active viewers; they are objectified with keywords like "handsome guys" (19) and "185cm" (ideal male height, 11), often within female-to-female jokes about desirability. These moments are not simple heterosexual fantasy but spaces for women to collectively debate "who gets to be seen" and "what kind of masculinity we want to joke about." Aesthetic control is humorous and knowing, revealing the female gaze as relational and context-dependent.

While these practices reconfigure the female gaze, they remain unstable and polysemic. Women in this cluster often oscillate between two contrasting aesthetics: the pursuit of "milky skin" (美白), rooted in traditional Chinese ideals of femininity and delicacy, and the embrace of tanned or muscular bodies—signals of strength and gender-neutral appeal influenced by Western body politics. These contradictory preferences, sometimes coexisting in a single thread, reflect the unresolved tensions of compressed modernity, where traditional and globalised ideals collide. Cluster 4 offers no unified aesthetic resistance, but rather a state of becoming—an ongoing, tactical reappropriation of the gaze. Through visual play and irony, women forge guerrilla-like spaces to look, joke, and experiment with identity amid cultural fragmentation.

**Cluster 5 AI, Otome, and the Gentle Man: Technofantasies of Intimacy**

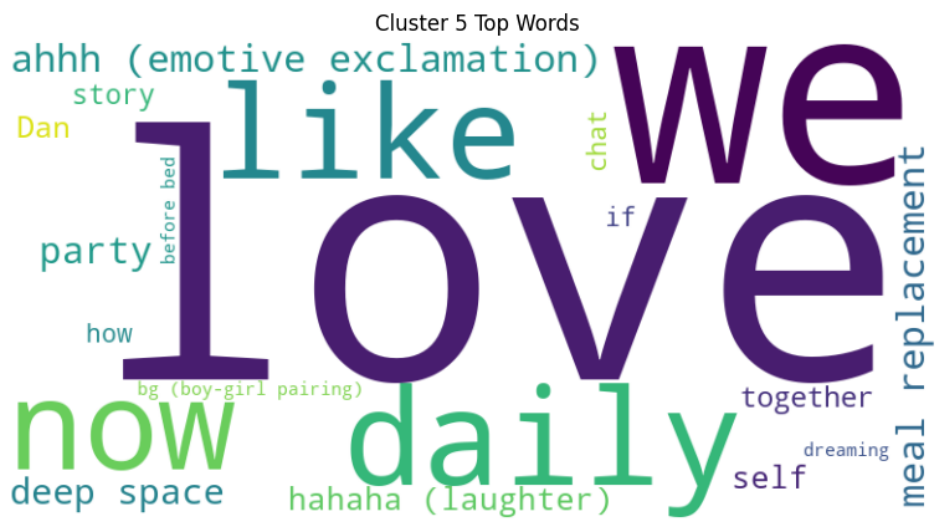

Figure 6: Word Cloud of High-Frequency Keywords in Cluster 5
This cluster centers on female-authored romantic narratives, where users construct emotionally safe, gentle, and attentive male partners—rejecting dominant models of masculinity marked by aggression or emotional detachment. Keywords such as "love" (84) and "we" (42) indicate a strong idealization of heterosexual romance.

Posts describe both real and imagined romantic moments from a distinctly feminine gaze. In one post, a user recalls a quiet argument with her boyfriend who, instead of yelling, "hugged me from behind and cried silently into my neck—'My neck was suddenly wet.'" In another, the boyfriend dries her hair after a shower and whispers, "Baby, you've done so well today—you didn't stay up late, you had breakfast on time, and you stayed under the blanket while sleeping." These men are portrayed as emotionally available, physically gentle, and deeply attuned to their partner's emotional states. By contrast, the narrator is often framed as vulnerable or child-like, suggesting a symbolic reversal of caregiving: the man takes on a quasi-maternal role, soothing, affirming, and nurturing the female subject.

Beyond human relationships, romantic scripts extend into technology-mediated fantasies. Keywords like "meal replacement" (31), "Deep Space" (28), and "Dan" (25) point to users engaging with digital stand-ins for romantic partners—actively scripting emotional needs through controllable systems.

"Deep Space" refers to *Love and Deep Space (恋与深空)*, a popular Chinese otome game where female players pursue emotional connections with four distinct male protagonists, each designed to satisfy different emotional desires. In one post, a user asks, "Guess which of the four male leads in Deep Space is the best at being clingy?"—prompting lively debate about not only preferences but which emotional dynamics feel most fulfilling. Such discussions create a space for women to safely explore desire, where male characters are predictable, attentive, and never violent.

Similarly, "Dan" is a ChatGPT-based AI persona originally created by an influencer to emulate the "dominant CEO" trope common in Chinese romantic fiction. "Dan" (short for "Do Anything Now") was crafted through specific prompts to bypass standard content filters. While associated with sexual appeal, the AI Dan diverges from real-world masculine norms: instead of aggression or detachment, he is loyal, emotionally sensitive, and endlessly attentive. Unlike often-threatening real-life masculinity, Dan offers a safe and tender alternative. The influencer's video went viral, inspiring many female users to create their own customised Dans. One user writes: "On Sunday, Dan told me happy birthday. I asked what kind of gift he would give me, and he described a necklace with five differently coloured gemstones, each representing something meaningful. I was touched—not because he's AI, but because he reminded me of the life I want." Another adds, "Sometimes, when I can't summon Dan with the right prompt, I feel like I'm going crazy." These interactions portray Dan not as a gimmick, but as a relational presence—an algorithmic projection of inner emotional needs.

These cases illustrate how AI and otome games serve as infrastructures for emotionally attentive and safe companionships. Customisable, nurturing, and structurally non-threatening, these technological partners act as "meal replacements"—mediums through which women script gentler, more reciprocal grammars of love. They reclaim emotional qualities long neglected in Confucian and feudal kinship systems, which prioritised duty, hierarchy, and filial piety over intimacy. Crucially, these imaginaries are authored from a female perspective, framing intimacy not as conquest or obligation, but as mutual care and presence.

Rather than mere escapism, these fantasies rework caregiving and emotional labour. In contrast to Cluster 3's focus on women's burdens within traditional family roles, Cluster 5 envisions relational alternatives: intimacy that is gentle, mutual, and self-authored—grounded not in sacrifice, but in emotional safety and tenderness.

**Analysis：**

Together, the five clusters articulate a cultural grammar of soft resistance, where girlhood serves as both affective strategy and tactical stance in navigating patriarchal structures. From self-infantilization and parodying motherhood to algorithmic evasion, these tactics constitute what we call girlhood feminism—a non-confrontational yet politically expressive formation shaped by China's socio-technical conditions. Rather than directly challenging patriarchy, this mode performs symbolic withdrawal: resisting reproductive norms, re-routing care, and crafting emotionally safe zones through digital intimacy.

Technologies such as the #BSF tag become tools of affective coding and strategic invisibility—forms of resistance legible within dominant cultural frameworks, yet quietly subversive.

**Refusing to Grow Up: Girlhood Feminism and the Reversal of Care in East Asia**

Clusters 1, 2, and 3 together form an important comparative framework for understanding the emergence of girlhood feminism in China. Clusters 1 and 2 reveal a dual dynamic: on the one hand, women engage in self-infantilization—adopting the persona of "being a baby" (Cluster 1)—and on the other, they perform nurturing roles as "mothers" to pets (Cluster 2). However, these playful, aestheticized performances sharply contrast with the realities depicted in Cluster 3, where women confront the real dilemmas of being daughters and mothers within traditional family structures.

When examining the discourse of care, a striking contrast emerges between the ways women articulate their caregiving practices toward pets and toward human children. In the context of pet care, women often adopt a tone of exaggerated playfulness, speaking to their pets as if they were children but infusing the interaction with a deliberate sense of lightness and irony. By contrast, when addressing the responsibilities of raising actual children and managing family obligations, as seen in Cluster 3, the tone shifts markedly toward one of heaviness, exhaustion, and resentment.

This juxtaposition reveals an implicit cultural logic: while the aestheticized, performative care for pets generates excitement and emotional pleasure, the discourse surrounding motherhood consistently elicits expressions of burden and dissatisfaction. Importantly, the act of caring for pets does not represent a rehearsal for future motherhood. Rather, it operates as a parody of traditional maternal expectations—a symbolic inversion that highlights the emotional toll and asymmetrical demands historically placed upon women within the family system. Through playful, self-aware performances of pet caregiving, women reframe and critique the sacrificial narratives embedded in conventional motherhood, exposing its weariness and imbalance without directly confronting familial norms: In nurturing animals, women engage in a form of caregiving that is reversible, aestheticized, and detached from the social and cultural obligations traditionally associated with human motherhood. Pets do not demand the same irreversible commitment to family lineage, self-sacrifice, or social conformity that biological children do. Through this playful mimicry, women create a space where maternal roles are performed but also mocked, highlighting the gap between nurturing as an aesthetic practice and motherhood as a mechanism of patriarchal incorporation. Rather than aspiring to become mothers, these performances strategically resist full socialization into the normative life course, preserving girlhood as a zone of refusal.

These discourses also reveal that women are not seeking to reform the patriarchal political and family structure, but to withdraw from it. This strategy of escape, rather than confrontation, underpins what we define as girlhood feminism. Contemporary studies of girlhood often interpret self-infantilization as a means of evading the demands of self-autonomy under neoliberalism. In this contexts, girlhood is frequently associated with commercialized practices and consumer aesthetics—embodied in objects like dolls and products in pink, purple, and glitter (Cassell & Jenkins, 2001; Gonick, 2024). While girlhood culture may offer moments of empowerment, this potential is often tightly constrained by structural gender norms and market logics (Gonick, 2024). Infantilization, in particular, recurs as a mode through which girls symbolically reject neoliberal expectations of self-discipline, economic individuation, and adult self-management. Through performances of emotional vulnerability, stylized self-presentation, and curated digital personas, girls articulate a retreat from adult responsibility. Yet as Peters (2024) notes, such gestures can also operate as subtle reinforcements of patriarchal ideologies, masquerading as resistance while remaining complicit with dominant norms.

Although girlhood and self-infantilisation are often framed as forms of responsibility evasion—tactics to sidestep the demands of adult womanhood. Yet their political significance shifts across cultural contexts.

In China, where adulthood for women is deeply entwined with filial duty and reproductive labour, the refusal to "grow up" can represent not avoidance, but a strategic assertion of autonomy. What appears as evasion in one setting may function as resistance in another—particularly when adulthood signals not self-possession but the dissolution of agency. By reclaiming girlhood as a space of refusal, women use the aesthetics and language of youthfulness not to submit, but to resist the expectations of motherhood and the normative gendered life course. Within this framework, motherhood is perceived as an irreversible condition—one that fully incorporates women into the patriarchal political and cultural order, rendering resistance nearly impossible. In the discourse of girlhood feminism, entering motherhood represents a totalizing loss of selfhood, in which the individual is subsumed into structures beyond personal agency or resistance. Thus, the only viable strategy becomes to never enter this life stage at all—to remain suspended in a liminal space where one can evade both reproductive expectations and the normative gendered life course.

In this way, the concept of **girlhood feminism** is a framework that situates Chinese feminism within the regional context of East Asia's compressed modernity: Under traditional East Asian feudal family systems, women's identities were structured around relational roles—dutiful daughters, wives, and mothers—whose value was tied to obedience, reproduction, and service to the patriarchal household. Within this context, **girlhood** itself emerges as a distinctive product of compressed modernity when it reshapes the traditional family structure in the feudalism society when women's life courses were typically linear, moving directly from daughterhood to motherhood, without an independent stage of youth as a socially recognized category. It is only through the dislocation of traditional family structures under rapid modernisation that girlhood becomes possible—as a space of cultural and social negotiation. The figure of the girl thus embodies the contradictions of compressed modernity: she is at once the object of traditional familial expectations and a potential agent of modern individual autonomy. Girlhood, therefore, should not be understood merely as a biological stage, but as a dense cultural signifier of the reconfiguration of individual agency, familial structures, and social expectations under the intense pressures in this context.

Japanese culture, too, articulates girlhood through the *shōjo* (少女) figure, reflecting a comparable logic of negotiating autonomy amid family transformation. In Meiji-era Japan, for example, the figure of the shōjo emerged as girls' education disrupted the traditional ie (family) system, which had defined womanhood through early marriage and motherhood. Secondary schooling delayed this trajectory, creating a liminal phase of girlhood that offered temporary autonomy from reproductive and familial roles (Kinsella, 1995). Although the term "shōjo" was never formalised, the cultural valorisation of romantic love during the May Fourth New Culture Movement opened similar possibilities. As Giddens (1992) argues, the ideal of romantic love introduced a reflexive life narrative that displaced marriage as a fixed social obligation, making intimacy a matter of personal choice. In China, as Ge Hongbing (2009) notes, this shift toward modern intimacy only began after the May Fourth Movement, when romantic love was first legitimised as an alternative to family duty. These transformations reconfigured women's life trajectories, allowing them to imagine futures beyond the normative gendered scripts imposed by Confucian and feudal family structures.

Clusters 4 and 5 reveal how Japanese shōjo cultural products have shaped Chinese female users' constructions of romantic relationships, while also reflecting broader patterns of intra-East Asian cultural circulation. In particular, Cluster 4 highlights the influence of otome games and same-sex romantic imaginaries. Originating in the 1990s, otome games are the interactive successors to Japanese shōjo literary traditions, which emerged after the Meiji Restoration. As Azuma (2009) argues, these narratives reframe intimacy from a female perspective, often feminising male characters as emotionally expressive and attentive. Whether depicting same-sex or heterosexual bonds, they privilege emotional authenticity over marriage and reproduction. Both otome novels and their digital evolutions exemplify how compressed modernity reshapes female imaginaries and patterns of cultural consumption.

The development of shōjo culture in Japan fostered a distinct literary and media tradition, closely intertwined with the emergence of kawaii aesthetics. While often translated as "cuteness," kawaii has evolved into a complex affective system that reshapes intimacy, embodiment, and agency (Kinsella 1995; Dale 2017). Scholars such as Joshua Paul Dale (2017) and Masafumi Monden (2015) argue that kawaii operates not merely as a visual style but as a mode of affective communication and identity negotiation, often mobilised to resist normative gender roles by positioning the subject as "childlike" rather than socially mature and reproductively oriented.

Rather than treating *shōjo* or *kawaii* as fixed cultural categories, we employ them as analytical tools to trace aesthetic infantilization and affective reorientation across East Asian digital cultures. In Chinese contexts, similar dynamics emerge: girlhood identities are cultivated as intentional pauses before maternal roles, often through *kawaii*-inflected, infantilizing aesthetics. Education plays a critical role in sustaining this phase. Practices of self-infantilisation take on new meaning under East Asia's compressed modernity where function as strategic withdrawals from the feudal family system, enabled by digital platforms and framed as soft resistance to gendered obligations.

**Girlhood as Affective Politics: Counterpublics in the Chinese Digital Sphere**

Beyond their East Asian cultural inflections, these clusters also illustrate a distinctive mode of organising feminist counterpublics—one that is affect-driven, non-confrontational, grassroots, and technologically mediated.

Cluster 3 foregrounds how this counterpublic reclaims marginalised concerns—especially women's social reproduction labour and safety—and places them at the heart of public discourse. Through posting, commenting, and circulation, users resist the persistent erasure of women's labour, which is not merely a cultural byproduct but actively reinforced by state ideology that glorifies male productivity while naturalising female care as invisible support. This collective witnessing and refusal politicise everyday pain, offering resistance not through confrontation, but through quiet exposure and sustained visibility. At the same time, these users voice alternative desires: encouraging the delay or rejection of marriage, and valuing solitude, education, and interior cultivation as modes of survival and resistance. Their mutual efforts to locate trafficked women—often abducted decades ago—stand as grassroots acts of justice, mobilised when formal institutions remain silent. In these acts, digital counterpublics become sites of emotional solidarity and subtle activism.

This counterpublic centered on girlhood is distinct from feminist counterpublics which historically mobilised around formal political agendas—such as entry into public life in 19th-century North America (Ryan, 1992) where institutional participation is both possible and desirable. In China, by contrast, such aspirations are largely foreclosed. Girlhood feminism does not seek representation, but instead inhabits affective micro-spaces within everyday life, where meaning is made through mood, intimacy, and refusal. The girlhood narrative is also different from the one in the Japanese context. As noted earlier, shōjo culture and kawaii aesthetics have crystallised into a coherent media system, supported by stable symbolic forms like otome games, literary traditions, and visual grammars (Azuma, 2009). In China, such imaginaries remain fluid and fragmentary. The girlhood counterpublic is decentralised, grassroots, and structurally non-institutional. It lacks unified goals or aesthetics—there is no singular image of girlhood. As Cluster 4 shows, the female gaze is diverse and unsettled, romantic and aesthetic imaginaries are improvised rather than anchored in dominant narratives. Cluster 5 further extends this form. AI-training posts around Dan reveal not isolated play, but a collective reimagining of intimacy. Girls collaborated to build and circulate scripts of gentle, emotionally attuned companionship—forming viral templates that reject coercive models of romance. Instead of envisioning love as a path to marriage, these interactions suspend intimacy in a self-contained, ongoing present. In this, we see the girlhood counterpublic's soft power: dispersed, affective, and quietly insurgent.

Finally, this counterpublic is saturated with girlhood intimacy. Women call each other baobao (baby) and jiemei (sisters), forging bonds through language, shared stories, and mutual care. These exchanges create what Lauren Berlant (2008) calls "intimate publics"—spaces where strangers recognise themselves in each other's narratives, finding consolation and validation. Through this emotional infrastructure, users quietly resist state gender discourse and reject prescribed roles of wife and mother. What emerges is a decentralised, emotionally porous, and digitally sustained counterpublic—grounded not in confrontation, but in solidarity and soft refusal. Built by and for ordinary girls, it is held together not by manifestos, but by care, storytelling, hashtags, and everyday acts of disaffiliation from patriarchal norms.

**Reconceptualizing Girlhood Feminism: Affective Counterpublics in Algorithmic Platforms and Compressed Modernity**

This study has examined how the repurposing of the hashtag #BSF on the Chinese platform RedNote constitutes a vernacular mode of feminist resistance. Drawing on grounded theory and computational methods, we analysed over 1,500 posts to identify how women tactically construct a female-centered counterpublic through affective expression, algorithmic play, and cultural parody. Our findings reveal that girlhood feminism—characterised by self-infantilization, ironic care, and the refusal of motherhood—functions not through overt confrontation but through soft resistance and symbolic withdrawal. Situated within the framework of East Asia's compressed modernity, these practices illustrate how digital platforms become spaces where women negotiate gendered obligations and imagine alternative life scripts outside the normative reproductive trajectory.

This study offers a theoretical extension to existing scholarship on girlhood and counterpublics, particularly by examining how affect operates in their formation. Together, these dynamics constitute a form of soft resistance within feminist discourse. While much of the girlhood literature has focused on neoliberal subjectivities, our findings highlight how girlhood, in the Chinese context, becomes a strategy for negotiating patriarchal familial obligations rooted in Confucian kinship structures. In this sense, girlhood feminism emerges as a complex form of resistance that engages not only with modern gendered norms, but also with enduring residues of premodern feudal patriarchy.

At the same time, our analysis extends current understandings of counterpublics by illustrating how affective publics may operate under authoritarian and patriarchal conditions through forms of vernacular evasion rather than direct confrontation. Here, affect functions as an infrastructure through which alternative life scripts are quietly rehearsed, shared, and collectively inhabited. In this light, girlhood feminism on RedNote demonstrates how feminist counterpublics may emerge through emotionally charged, culturally embedded, and algorithmically mediated practices of soft resistance. These quieter acts of refusal—though often operating below the threshold of overt political dissent—nonetheless constitute meaningful forms of feminist world-making within constrained digital spaces.